\documentclass{czjphys}         
\usepackage{graphicx}
\newcommand{\lsim}{\,{\buildrel < \over {_\sim}}\,}
\newcommand{\gsim}{\,{\buildrel > \over {_\sim}}\,}
\newcommand{\sqrtsNN}{\sqrt{s_{\scriptscriptstyle{{\rm NN}}}}}

\newcommand{\gev}{\mathrm{GeV}}
\newcommand{\tev}{\mathrm{TeV}}

\newcommand{\mum}{\mathrm{\mu m}}

\newcommand{\PbPb}{\mbox{Pb--Pb}}

\newcommand{\pt}{p_{\rm t}}

\newcommand{\dEdx}{{\rm d}E/{\rm d}x}

\newcommand{\Dz}{\mbox{$\mathrm {D^0}$}}

\newcommand{\Jpsi} {\mbox{J\kern-0.05em /\kern-0.05em$\psi$}\xspace}

\begin{document}
\title{\mbox{Measuring beauty production in Pb--Pb collisions at the LHC}
       via single electrons in ALICE}  
%
\authori{\underline{A.~Dainese}, F.~Antinori, C.~Bombonati, M.~Lunardon, and R.~Turrisi, for the ALICE Collaboration}      \addressi{Dipartimento di Fisica ``G.~Galilei'', Universit\`a degli Studi di Padova, and INFN, Italy}
\authorii{}     \addressii{}
\authoriii{}    \addressiii{}
\authoriv{}     \addressiv{}
\authorv{}      \addressv{}
\authorvi{}     \addressvi{}
%
\headauthor{A.~Dainese et al.}            
\headtitle{Measuring beauty production via single electrons in ALICE} 
\lastevenhead{A.~Dainese et al.:  Measuring beauty production via single electrons in ALICE} 
%
\pacs{25.75.-q, 14.65.Dw, 13.25.Ft}     
\keywords{heavy-ion collisions, beauty hadrons, semi-leptonic decays} 
\refnum{A}
\daterec{October 28, 2005}    
\issuenumber{0}  \year{2005}
\setcounter{page}{1}
\maketitle

\begin{abstract}
We present the expected ALICE performance
for the measurement of the $\pt$-differential 
cross section of electrons from beauty decays in 
central Pb--Pb collisions at the LHC. 
\end{abstract}

\section{Introduction}

The ALICE experiment~\cite{alicePPRv1} will study proton--proton, 
proton--nucleus and nucleus--nucleus collisions at the LHC, 
with centre-of-mass energies per nucleon--nucleon (NN) pair, $\sqrtsNN$,
of $5.5~\tev$ for \PbPb, $8.8~\tev$ for \mbox{p--Pb}, and $14~\tev$ for pp. 
The primary physics goal of ALICE is the study of 
the properties of QCD matter at the energy densities of several hundred 
times the density of atomic nuclei that will be reached in central 
\PbPb~collisions. Under these conditions
a deconfined state of quarks and gluons 
is expected to be formed.
Heavy quarks, and hard partons in general, would probe this medium
via the mechanism of QCD energy loss. In particular, measuring the
high-$\pt$ suppression of beauty hadrons and comparing it to that of 
light-flavour and charm hadrons, would allow us to investigate 
the predicted quark-mass dependence of parton energy loss~\cite{prd}.
Besides energy loss studies, the measurement of the beauty 
cross section provides the normalization for quarkonia production, needed
to address
medium effects on quarkonia.

\section{Expected rates and detection strategy}

Beauty decays with an electron in the final state 
have a `global' branching ratio of $\approx 21\%$:
$\approx 11\%$ for direct semi-electronic decays, ${\rm B}\to e\nu+X$, 
and $\approx 10\%$ for semi-electronic decays via a charm hadron, 
${\rm B\to D}(\to e\nu+X)+X'$~\cite{pdg}. 
The heavy-flavour production yields in Pb--Pb 
collisions, used as a baseline for the ALICE
performance studies, are obtained~\cite{notehvq} by scaling,
according to the number of binary collisions, the results of 
QCD calculations at 
next-to-leading order accuracy for pp collisions~\cite{mnr}.
The expected yields of beauty hadrons
and beauty-decay electrons in a 0--5\% central Pb--Pb collision at 
$\sqrtsNN=5.5~\tev$ are about 9.0 and 1.9, respectively, of which 
about 1/4 within the ALICE central barrel acceptance $|\eta|<0.9$.

As in central \mbox{Pb--Pb} collisions at the LHC between 3000 and 10\,000
charged hadrons may be produced in the region $|\eta|<0.9$, high-performance 
electron identification will be necessary.
In ALICE, a combined selection on 
$\dEdx$, measured in the Time Projection
Chamber (TPC), and transition
radiation, measured in the Transition Radiation Detector (TRD),
is expected reject 99.99\% of the pions ($10^{-4}$ misidentification 
probability)
and 100\% of the heavier hadrons, 
while correctly tagging 80\% of the electrons.

The main sources of background electrons are: 
decays of primary D mesons, which have a branching ratio 
of $\approx10\%$ in the semi-electronic channels~\cite{pdg}, and
an expected production yield larger by a factor about 20 
with respect to that of B mesons~\cite{notehvq}; 
$\pi^0$ Dalitz decays and 
decays of light mesons (mainly $\rho$, $\omega$, K);
conversions of photons in the beam pipe or in the inner layer of the
Inner Tracking System (ITS), and pions misidentified as electrons.

Given that electrons from beauty have average transverse
impact parameter (distance of closest approach to the interaction 
vertex in the plane transverse to the beam direction) $d_0\simeq 500~\mum$, 
it is possible to 
minimize the background contributions by selecting 
electron tracks displaced from the primary vertex. 
Tracking in the TPC and 
ITS, that includes two layers of silicon pixel detectors, 
provides a measurement of the 
impact parameter with a resolution 
$\sigma_{d_0}\,[\mum]\approx 11+53/(\pt\,[\gev/c])$, i.e.\,better 
than $65~\mum$ for $\pt>1~\gev/c$. The $d_0$ distributions reported 
in Fig.~\ref{fig:d0andSTAT} (left) show that 
a lower cut on the impact parameter
allows to reject misidentified $\pi^\pm$ and $e^{\pm}$
from Dalitz decays and photon conversions 
(the former mostly come from the primary vertex and 
the latter have small impact parameter for $\pt\gsim 1~\gev/c$).

\section{Results: sensitivity on the b-decay electron cross section}      

\begin{figure}[!t]
  \begin{center}
    \includegraphics[width=.48\textwidth]{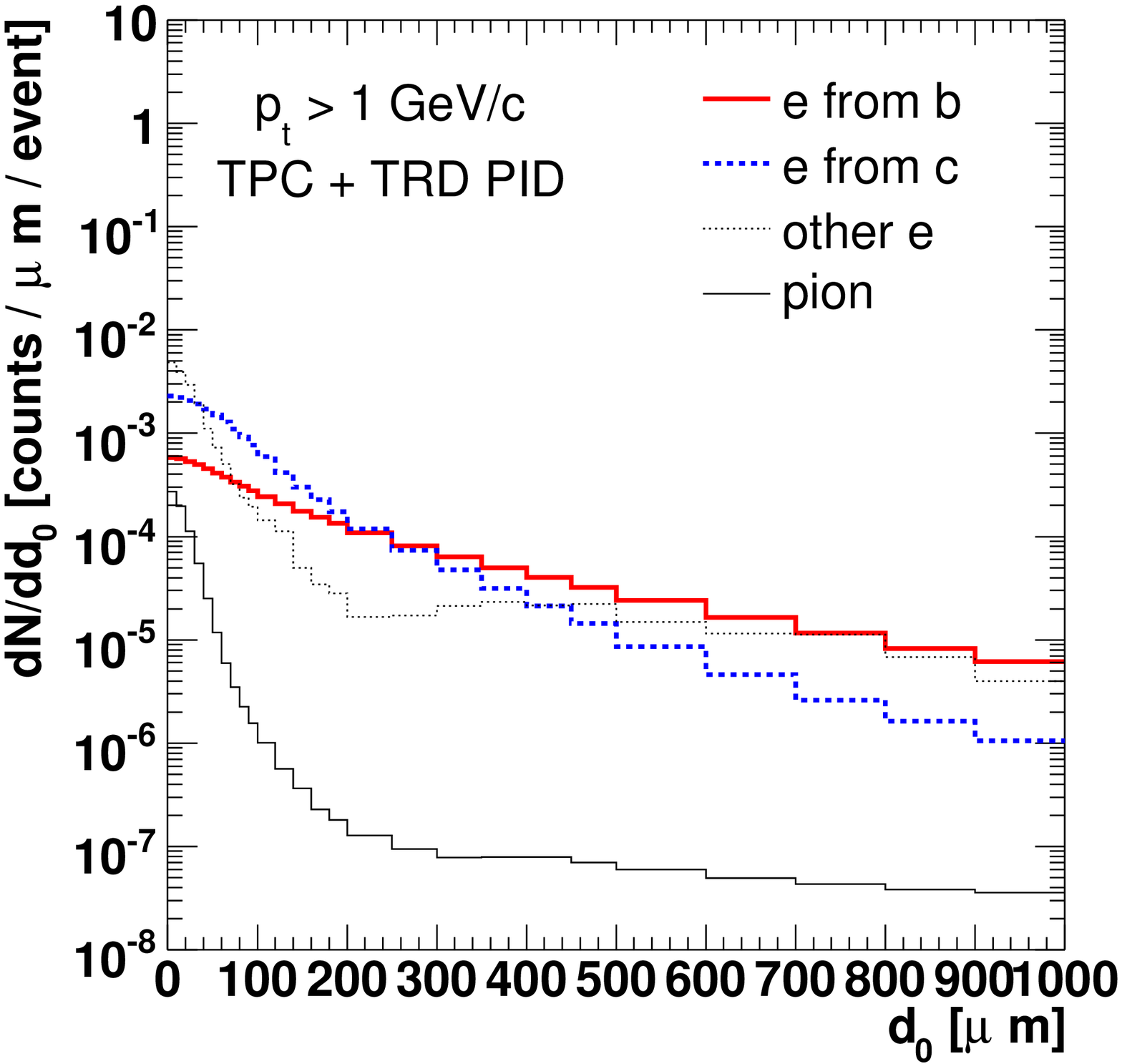}
    \includegraphics[width=.51\textwidth]{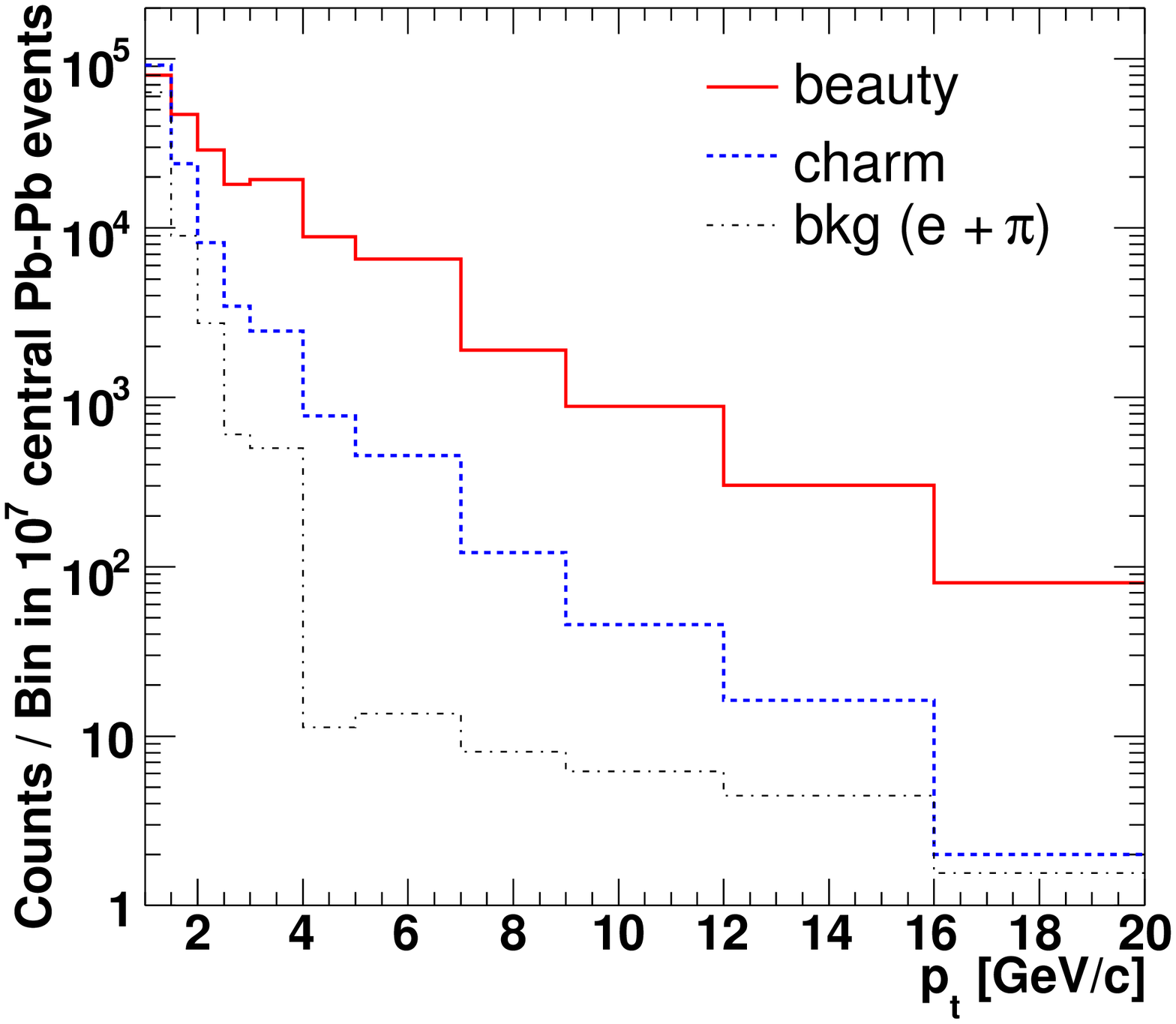}
    \caption{Left: impact parameter distribution for identified electrons from 
             beauty decays, from charm decays, 
             identified electrons from the background and 
             pions tagged as electrons. Combined TPC--TRD identification is
             applied.
             Right: expected statistics in counts/$\pt$-bin for $10^7$ central 
             Pb--Pb events (one month of data taking) 
             for electrons from beauty, from charm and for
             the background 
             (misidentified pions or electrons from other sources),
             with the cut $d_0>200~\mum$.}
    \label{fig:d0andSTAT}
  \end{center}
\end{figure}

As it can be seen from Fig.~\ref{fig:d0andSTAT} (left),
the impact parameter cut $d_0>200~\mum$ is expected to  
give a b/c ratio larger than unity for the electron sample
with $\pt>1~\gev/c$. 
This cut also removes most of the background 
electrons from $\pi^0$ Dalitz decays and $\gamma$ conversions, as
well as the primary $\pi^\pm$ misidentified as electrons.

The expected statistics
for electrons from b decays for $10^7$ central \PbPb~events (one month 
of data taking at nominal Pb--Pb luminosity) is shown in the right-hand panel
of Fig.~\ref{fig:d0andSTAT}, along with the residual background 
contributions. 
About $8\times 10^4$ beauty electrons are expected above $2~\gev/c$ in 
$\pt$,
allowing the measurement of the $\pt$-dif\-fe\-ren\-tial 
cross section in the range $2<\pt<18~\gev/c$. 
The residual contamination of about 10\% 
of electrons from prompt charm decays, from misidentified charged pions
and $\gamma$-conversion electrons is accumulated in the low-$\pt$ region.

After applying the $d_0$ cut, in a given $\pt$ interval [e.g.\,the bins in 
Fig~\ref{fig:d0andSTAT} (right)], $N$ `electrons' will be counted, 
being the sum of different contributions, $N=N_{\rm b}\,({\rm beauty})+N_{\rm c}\,({\rm charm})+N_{\rm bkg}\,({\rm bkg}~e~{\rm and~misid.~\pi})$.
The following procedure is foreseen 
to extract the cross section of electrons from beauty.
\begin{enumerate}
\item
Subtraction of charm decay electrons. We plan to use the 
 cross section for $\Dz$ mesons, measured in ALICE in the $\rm K^-\pi^+$ 
decay channel~\cite{D0jpg,thesis}, to estimate the $N_{\rm c}$ contribution.
This will introduce a systematic error coming from the statistical and
systematic errors on the $\Dz$ cross section. The error is expected to 
be smaller than 5\% for $\pt>2~\gev/c$.
\item
Subtraction of the remaining background electrons and misidentified 
pions, on the basis of Monte Carlo simulations tuned on the measured 
light-flavour hadron production. This subtraction also introduces a 
systematic error, which we currently assume to be small, i.e.\,negligible 
with respect to other error sources.
\item
Correction of the number of beauty electrons, 
$N_{\rm b}=N-N_{\rm c}-N_{\rm bkg}$,
for efficiency ($d_0$ cut, electron ID, tracking) and acceptance:
$N_{\rm b}^{\rm corr}=N_{\rm b}/\epsilon$. The correction will be done 
via Monte Carlo and we assume a systematic error of about 10\% on the 
correction factor $1/\epsilon$.
\item
Normalization of the corrected yield to a cross section for 
beauty electrons per binary collision in a given Pb--Pb centrality class
(e.g.\,0--5\%), $\sigma^{\rm e\,from\,b}_{\rm NN}$. For the 0--5\% class, 
a systematic error 
of about 9\% is expected to be introduced in the determination of the 
correction factor, which is proportional to the 
average number of binary collisions in the considered centrality 
class.
\end{enumerate}

\begin{figure}[!t]
  \begin{center}
    \includegraphics[width=.49\textwidth]{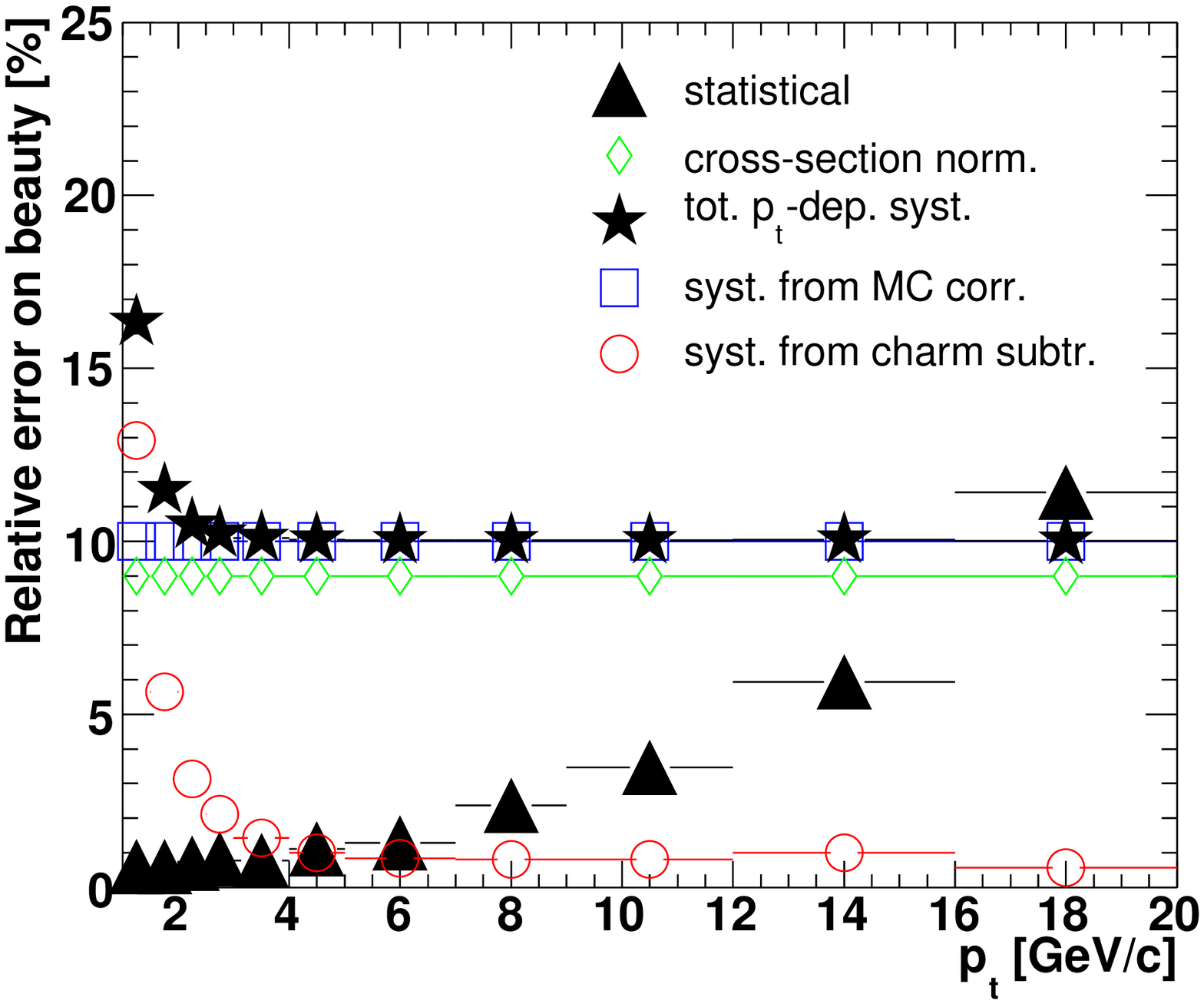}
    \includegraphics[width=.4\textwidth]{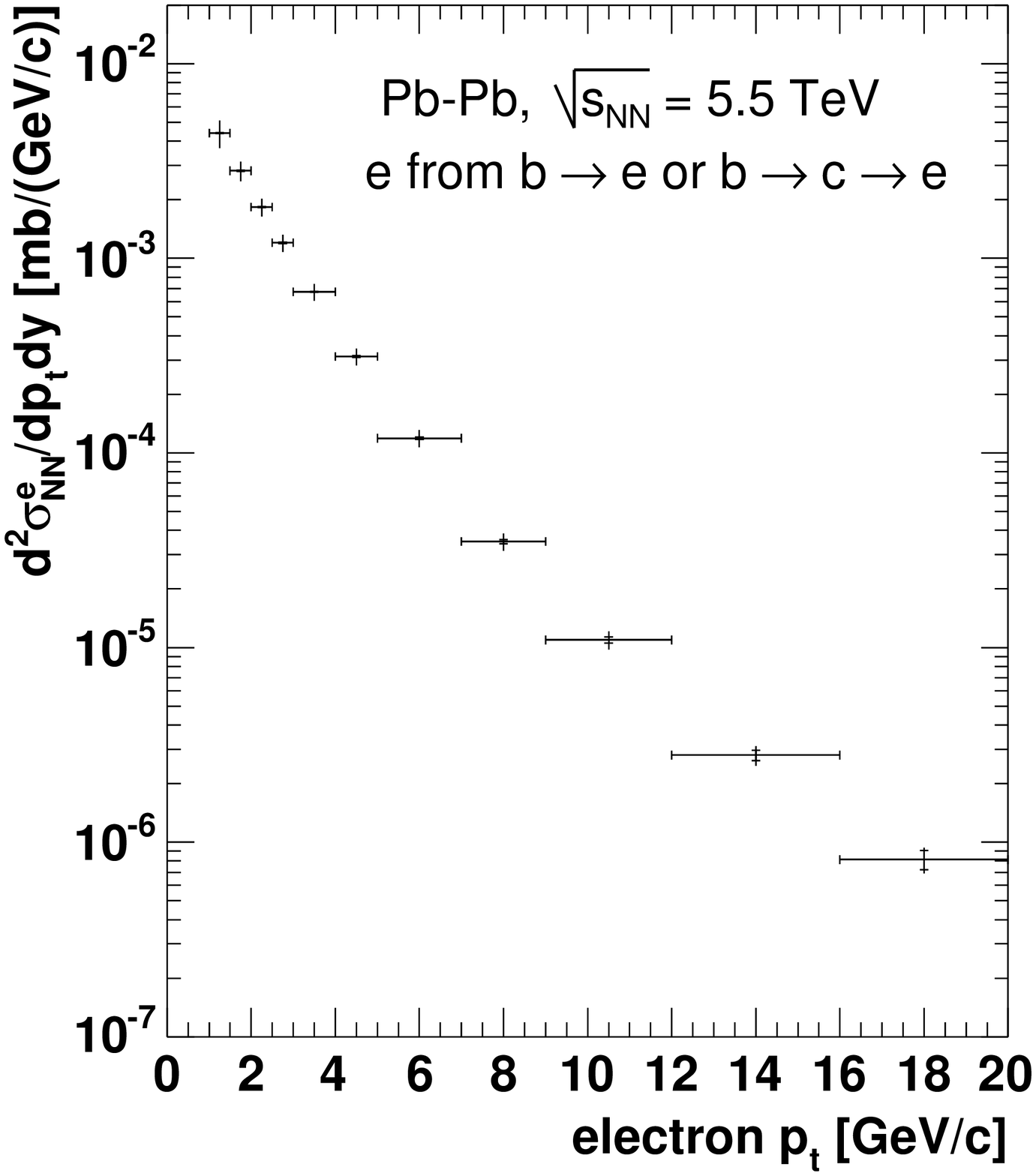}
    \caption{Left: error contributions for 
             electrons from b decays, relative to 
              $10^7$ central Pb--Pb events
             (one month of data-taking).
             Right: cross section per NN collision
             for electrons from b decays as a function of $\pt$, 
             as it is expected to be measured with $10^7$ central events;
             statistical errors (inner bars) and quadratic sum of statistical 
             and $\pt$-dependent systematic 
             errors (outer bars) are shown;
             the 9\% normalization error is not shown.}
    \label{fig:sigmae}
  \end{center}
\end{figure}

Figure~\ref{fig:sigmae} (left) shows a summary of the relative errors 
on beauty electrons, as a function of $\pt$. Besides the systematic
errors, we also show the relative statistical error 
$\sqrt{N_{\rm b}+N_{\rm c}+N_{\rm bkg}}/N_{\rm b}$: it is 
expected to increase from less than 1\% at $\pt=2~\gev/c$ to about 12\%
at $18~\gev/c$, for $10^7$ central events (one month at nominal luminosity).
The expected quality of the measured beauty electron cross section is 
shown in Fig.~\ref{fig:sigmae} (right). A procedure to extract,
starting from the $e$-level cross section, a B-level cross section 
(as a function of B $\pt^{\rm min}$) is under study.
Note that no medium-induced high-$\pt$ suppression is included
in the results shown in Fig.~\ref{fig:sigmae}; 
the predicted suppression of a factor 4--5~\cite{prd} 
would increase the relative 
statistical errors by a factor about 2. 

Preliminary simulation results on the ALICE capability to perform this 
measurement in pp collisions indicate that, combining Pb--Pb and pp data,
the nuclear modification 
factor of electrons from beauty can be obtained with good precision in 
the range $2\lsim\pt\lsim 20~\gev/c$. This would allow us to 
test the 
predicted mass-dependence of parton energy loss in the medium formed 
in Pb--Pb collisions.
        
\end{document}